\documentclass[lettersize,journal]{IEEEtran}
\usepackage{amsmath,amsfonts}
\usepackage{algorithm}%
\usepackage{algorithmic}
\usepackage{array}
\usepackage[caption=false,font=normalsize,labelfont=sf,textfont=sf]{subfig}
\usepackage{textcomp}
\usepackage{stfloats}
\usepackage{url}
\usepackage{verbatim}
\usepackage{graphicx}
\usepackage{booktabs}%
\usepackage{multirow}%

\def\BibTeX{{\rm B\kern-.05em{\sc i\kern-.025em b}\kern-.08em
    T\kern-.1667em\lower.7ex\hbox{E}\kern-.125emX}}
\usepackage{balance}

\begin{document}

\title{Robust Channel Learning for Large-Scale Radio Speaker Verification}

\author{
\IEEEauthorblockN{Wenhao Yang\IEEEauthorrefmark{1},
                  Jianguo Wei\IEEEauthorrefmark{1}, 
                  Wenhuan Lu\IEEEauthorrefmark{1},
                  Lei Li\IEEEauthorrefmark{2},
                  Xugang Lu\IEEEauthorrefmark{3}}  \\      
\IEEEauthorblockA{\IEEEauthorrefmark{1} College of Intelligence and Computing, Tianjin University, Tianjin, China}\\
\IEEEauthorblockA{\IEEEauthorrefmark{2} Department of Computer Science, University of Copenhagen, Copenhagen, Denmark}\\
\IEEEauthorblockA{\IEEEauthorrefmark{3} National Institute of Information and Communications Technology, Japan}\\
\IEEEauthorblockA{\{yangwenhao, jianguo\}@tju.edu.cn, lilei@di.ku.dk}

}


\maketitle

\begin{abstract}
    
Recent research in speaker verification has increasingly focused on achieving robust and reliable recognition under challenging channel conditions and noisy environments. Identifying speakers in radio communications is particularly difficult due to inherent limitations such as constrained bandwidth and pervasive noise interference. To address this issue, we present a Channel Robust Speaker Learning (CRSL) framework that enhances the robustness of the current speaker verification pipeline, considering data source, data augmentation, and the efficiency of model transfer processes. Our framework introduces an augmentation module that mitigates bandwidth variations in radio speech datasets by manipulating the bandwidth of training inputs. It also addresses unknown noise by introducing noise within the manifold space. Additionally, we propose an efficient fine-tuning method that reduces the need for extensive additional training time and large amounts of data. Moreover, we develop a toolkit for assembling a large-scale radio speech corpus and establish a benchmark specifically tailored for radio scenario speaker verification studies. Experimental results demonstrate that our proposed methodology effectively enhances performance and mitigates degradation caused by radio transmission in speaker verification tasks. The code will be available on Github\footnote{https://github.com/wenhao-yang/twowayradio}.
\end{abstract}

\begin{IEEEkeywords}
    Speaker Verification, Data Augmentation, Radio Communication, Channel Degradation
\end{IEEEkeywords}

\section{Introduction}

Speaker verification serves as an exemplary application of representation learning. In this task, a deep neural network classifier is initially trained on speech data to distinguish between speakers. Subsequently, the embedding layer and its preceding layers from this trained model are leveraged as a speaker vector extractor. This extractor generates speaker identity vectors, which embody the learned speaker representations. Beyond speaker verification, speaker embeddings have been applied to various speech processing applications, including speaker adaptation for speech recognition \cite{saon2013speaker}, voice conversion \cite{Lu2019OneShotVC}, speech synthesis \cite{jia2018transfer,zhang2021transfer}, and speech enhancement \cite{9054311}. Deep speaker verification has achieved remarkable success when trained with large-scale datasets like x-vectors \cite{snyder2018x} and ECAPA-TDNN \cite{desplanques2020ecapa} within the VoxCeleb \cite{NAGRANI2020101027} benchmark.

Speaker verification has nearly reached its performance ceiling in relatively simple scenarios, especially with the introduction of large-scale pre-trained models such as wav2vec2 \cite{baevski2020wav2vec}, HuBERT \cite{9585401} and WavLM \cite{chen2022wavlm}. High performance requires no serious signal degradation, similar channels, etc. Under this condition, speaker-independent or speaker-dependent variation is limited in those audios. Otherwise, when these conditions are not met, the performance of the algorithm may fall short of the practical application requirements \cite{fan2020cn,brown2021playing}. Therefore, evaluating and developing robust and reliable algorithms for complex speaker pronunciation, various speaking contents, and transmission channels remains an important research field in speaker verification \cite{gonzalez2014evaluating,qin2020hi,zheng20233d}. Among the contributing factors to performance degradation, channel discrepancies are one of the most significant elements. Channel discrepancies primarily arise due to differences in recording devices and signal transmission channels. Device variations encompass differences among distinct types of microphones employed in various devices for audio recording, such as high-fidelity microphones and Android and iOS smartphones \cite{bu2017aishell,qin2020interspeech}. Transmission channels include both wired and wireless methods, with wired telephony and Internet transmissions having been studied more extensively.

To enhance the robustness of speaker verification systems across various scenarios, a variety of techniques have been proposed. Mixing audio corpora from different channels or scenarios is one of the most effective methods, although it can lead to a linear increase in training time. Moreover, the need for training new models arises when data from new channels emerges. Apart from simple mixing, other approaches such as data augmentation \cite{snyder2018x,shahnawazuddin2020domain,wang2020investigation}, domain adaptation \cite{garcia2014supervised,lee2019coral+,lin2020multi}, adversarial training \cite{wang2018unsupervised,bhattacharya2019adapting,peri2020robust}, and transfer learning \cite{hong2017transfer,qin2019far,peng2023parameter} are actively pursued in robust speaker verification. 

However, research on speaker verification within radio communication environments remains relatively underexplored. This paucity of exploration can be attributed to two primary factors. Firstly, acquiring high-quality radio speech data suitable for speaker verification purposes poses significant challenges, as most existing datasets of this nature are neither freely accessible nor openly shared. Consequently, the establishment of an open benchmark for this scenario is imperative. Secondly, the process of speaker verification in radio contexts inherently encounters formidable obstacles, including substantial levels of noise interference and signal attenuation, which complicate speaker verification in such environments. Therefore, further investigation into current data augmentation methods for speaker verification under conditions of unknown noise is warranted. Our work addresses the challenge and proposes a robust pipeline for radio communication environments.

Based on these observations, this paper proposes a robust learning framework for speaker verification. Our contribution consists of the following three aspects:

\begin{enumerate}
\itemsep=0pt
\item We propose a Channel Robust Speaker Learning framework for enhancing the robustness of speaker deep neural networks effectively.
\item We demonstrate that combining specific augmentation methods with fine-tuning within this framework narrows the performance gap between standard audio and radio speaker verification, thereby enhancing reliability in typical scenarios.
\item We develop a radio transmission toolkit, compile a robust radio audio corpus, establish a benchmark for radio speaker verification, and empirically validate our approaches.
\end{enumerate} 

The rest of the paper is organized as follows: we present related works in Section.\ref{sec:related}. We introduce the methodology of our schema in Section.\ref{sec:meth}. Then, we carry out experiments, analyze the performance of state-of-the-art (SOTA) speaker verification systems on radio speech, explore to build robust systems, and discuss experimental results in Section.\ref{sec:experiments}. Finally, we summarize in Section.\ref{sec:conclusion}.

\section{Related Work}
\label{sec:related}

Speaker verification within radio communication is closely tied to robust learning and domain adaptation. Prior research has traditionally concentrated on more general scenarios rather than focusing on severely degraded conditions.

\subsection{Robust Learning}

Extensive research has been dedicated to developing robust speaker verification models by investigating several key areas. Data augmentation techniques, like adding noise and reverberation as in \cite{xie2019utterance}, manipulating children's speech speed and pitch in \cite{shahnawazuddin2020domain} or exploring SpecAugment in \cite{wang2020investigation}, generate realistic noisy inputs. Model architectures like ThinResNet \cite{xie2019utterance}, ECAPA-TDNN \cite{desplanques2020ecapa}, and CAM++ \cite{wangcam++} optimize parameter usage for enhanced performance. Loss functions, including AM-Softmax \cite{wang2018additive} and ArcFace \cite{deng2019arcface}, introduce additive margin for decision boundary in cosine similarity space. Score normalization methods, e.g., t-norm \cite{auckenthaler2000score} and As-norm \cite{matejka2017analysis}, minimize discrepancies between training and testing data by subtracting top-k similarity scores between those samples. While these strategies, when combined with large datasets and overparameterized models, significantly boost robustness, they often overlook specific scenarios like radio communication

\subsection{Domain Adaptation}

Channel degradation may lead to domain shifts or data distribution drift, impacting model performance. Substantial efforts have been dedicated to adapting models to various domains. Adversarial training, commonly employing the Gradient Reversal Layer and discriminative models to learn domain-invariant features, has emerged as a prevalent method in both speaker verification and other tasks \cite{ganin2016domain,wang2018unsupervised}. In \cite{yu17_interspeech}, a noise-robust bottleneck feature for speaker verification, generated by an adversarial network, was proposed. Another study proposed maximum mean discrepancy to adapt deep neural networks for new environments in speaker verification \cite{9054134}. Notably, Probabilistic Linear Discriminant Analysis (PLDA) and its variants, including CORAL combined with PLDA, offer a principled probabilistic framework for modeling speaker characteristics' variability and have demonstrated remarkable effectiveness \cite{ioffe2006probabilistic,lee2019coral+,wang2020generalized}. Furthermore, alternative approaches to enhance robustness and performance include leveraging multi-source data \cite{qin2019far}, adding multi-level adapters \cite{10446444}, applying meta-learning techniques \cite{zhang2021meta,lin2023model}, incorporating transfer learning \cite{hong2017transfer}, and utilizing large pre-trained models, like wav2vec2.0 \cite{baevski2020wav2vec} and WavLM \cite{chen2022wavlm}.

\subsection{Radio Signal Recognition}

Several studies have addressed radio speech, examining speech recognition, speaker verification, and related tasks under radio scenarios \cite{walker2012rats,plchot2013developing,o2016radio}. Researchers in \cite{plchot2013developing} transmitted audio samples using RATS and devised an adaptation method for robust speaker verification. In \cite{lin2020unified,trnka2021speaker}, similar to our research, audio samples were transmitted over the radio for secure airflow control systems. Text-dependent speaker verification in \cite{larcher2014extended} compared the effects of radio audio samples against clean ones. Unfortunately, these radio datasets are either not publicly available or difficult to access. Moreover, these studies did not fully consider the variability inherent in radio communication scenarios. In summary, with respect to speaker verification in radio environments, while comparable studies exist, they tend to be incomplete, and notably, the datasets involved in these studies are neither publicly available nor easy to reproduce.

\section{Methodology}
\label{sec:meth}
Considering speaker verification in radio scenarios, audio samples exhibit statistical distribution drift compared to clean data. Combining these two types of corpora for training a radio-robust model can significantly increase training time due to low sample efficiency. To address this issue, we adopt two modules: first, a data augmentation module to simulate the distribution of radio data, and second, a layer selection-based fine-tuning module for efficient training. Additionally, we introduce a data collection module to validate our methods, aimed at constructing datasets for optimizing parameters in both modules and benchmarking their performance.

\begin{figure*}[th]
	\centering
		\includegraphics[scale=1, width=0.85\linewidth]{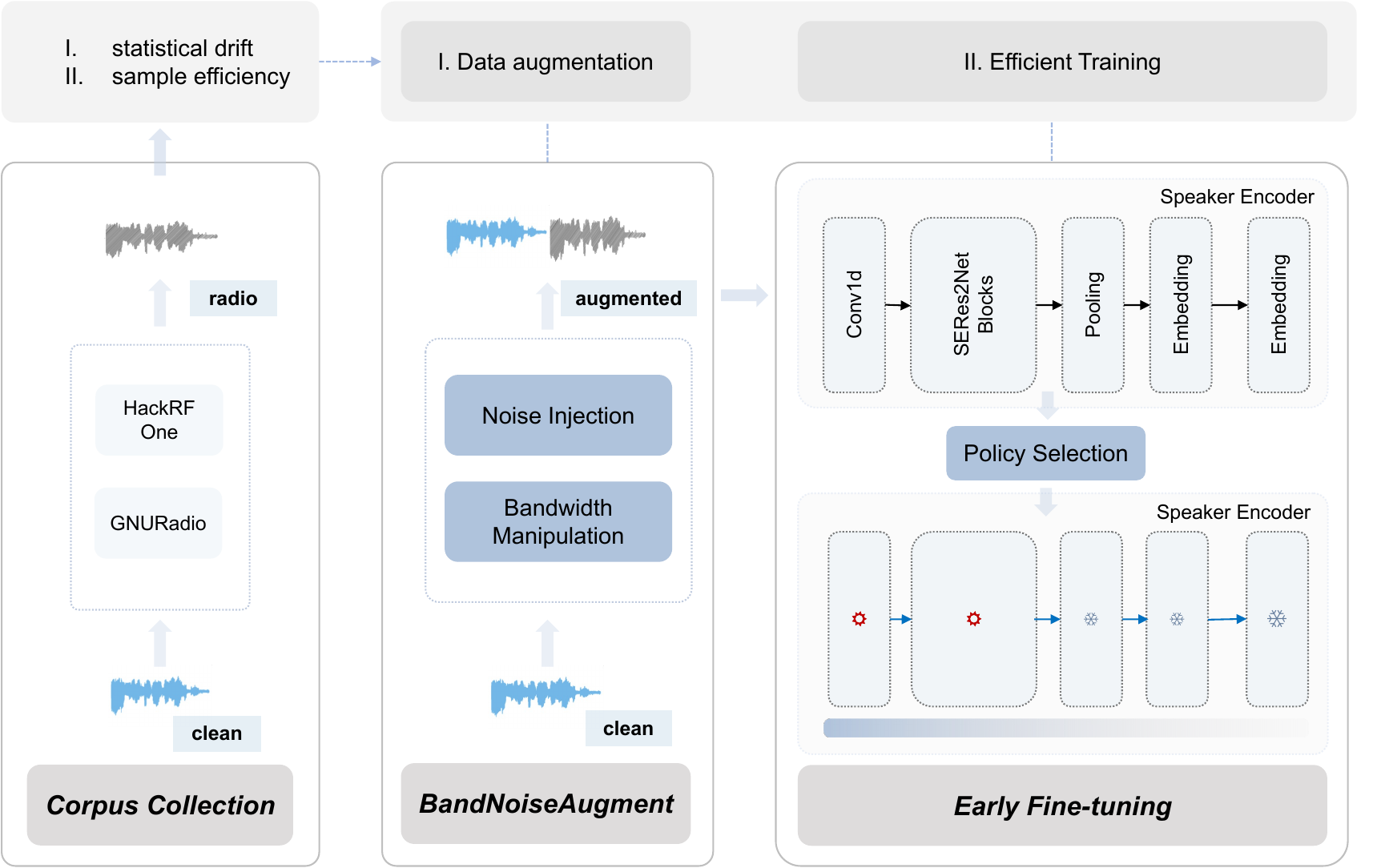}
	\caption{An overview of the Channel Robust Speaker Learning framework for speaker verification. Left: Corpus Collection for radio communication; Medium: BandNoiseAugment for audio corpus; Right: Fine-tuning for early-stage convolutional layers.}
	\label{FIG:framework}
\end{figure*}

The proposed Channel Robust Speaker Learning (CRSL) framework comprises three main components as in Figure.\ref{FIG:framework}: (i) collecting scenario-specific audio corpus, specifically radio audio; (ii) learning bandwidth-invariant representations using the BandNoiseAugment module; and (iii) fine-tuning early-stage convolutional layers with scenario-specific drift analysis.


\subsection{Corpus Collection}

\begin{figure}[h]
	\centering
		\includegraphics[scale=1, width=0.95\linewidth]{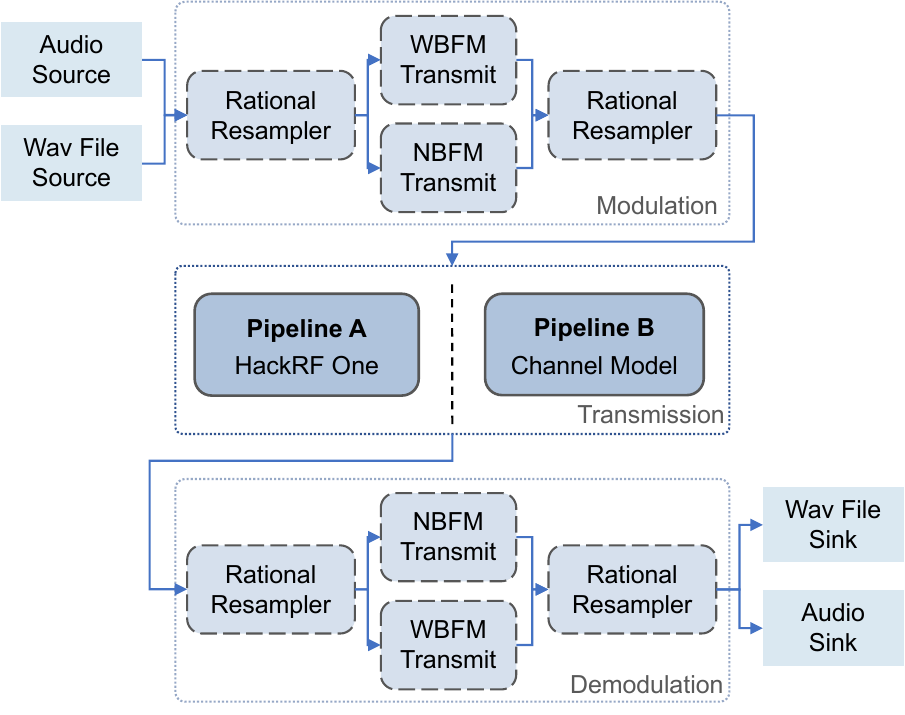}
	\caption{This is the GRC graph of the pipeline of the overall radio transmission framework. Pipeline A uses HackRF One to transmit the audio signal in the air. Pipeline B uses the Channel Model in GNU Radio to simulate the transmission with multiprocessing.}
	\label{FIG:pipeline}
\end{figure}

HackRF One is a Software Defined Radio (SDR) peripheral capable of transmission or reception of radio signals. We use HackRF One to simulate the process of Two-way Radio communication. GNU Radio \cite{blossom2004gnu} is a free \& open-source software development toolkit that provides signal processing blocks to implement software radios. In the development tool GNU Radio Companion (GRC), we use the GRC diagram to build the simulation of the radio communication data flow, as Figure.\ref{FIG:pipeline}. The radio corpus creation process can be outlined in the following steps:

\begin{enumerate}
\item \textbf{Loading and Preprocessing} Read audio WAV files from disk storage and resample to the specific sample rate of an NBFM or WBFM Transmitter. 
 
\item \textbf{Modulation and Transmission} Send the modulated signal to an Osmocom Sink Block. Configured with the designated HackRF One device ID, center frequency, and sampling rate, the block uses USB-connected hardware to broadcast the signal over the air.

\item \textbf{Reception and Amplification} Receive the signal using the Osmocom Source Block connected to the HackRF One device. Sample and amplify the received signal for subsequent processing stages.

\item \textbf{Demodulation and Saving} Pass the complex signal to an NBFM/WBFM Receiver Block for demodulation. Afterward, resample the demodulated signal and route it either to an Audio Sink for playback or a WAV file Sink for storage.

\end{enumerate}

Directly transmitting and compiling extensive datasets, such as VoxCeleb2, with over 2000 hours of audio, is time-consuming. Hence, we employed the GNU Radio toolkit to build a data collection pipeline capable of concurrent transmission of multiple samples. As in Figure.\ref{FIG:pipeline}, the radio corpus collection is organized into two pipelines: 

\textbf{Pipeline A:} Speech data is read using GNU Radio, with HackRF One handling the transmission and reception of radio signals. The Osmocom Sink and Osmocom Source blocks interface with the HackRF One device.

\textbf{Pipeline B:} Speech data is again processed using GNU Radio, but this pipeline emulates the sending and receiving of radio signals through a Channel Model block. The Channel Model incorporates an adjustable Noise Voltage parameter to simulate Additive White Gaussian Noise (AWGN) levels represented as a voltage. Pipeline B implements multiprocessing to address efficiency concerns in simulating airborne signal transmission, thus enabling efficient short-term simulation of a large volume of radio speech data transmissions.



\subsection{BandNoiseAugment} 

In the context of speech data transmission, given the constraints imposed by channel bandwidth, it is common to compress high-quality audio captured via microphone recordings to ensure real-time feasibility. This compression is typically achieved through downsampling and codec-based encoding, with compression ratios varying significantly across different scenarios. Current speaker verification frameworks inadequately account for this diversity in data augmentation strategies, often relying on time-frequency masking techniques, such as SpecAugment \cite{park2019specaugment} and EnvCorrupt \cite{ravanelli2021speechbrain}, to mimic information loss in audio datasets. In scenarios like radio communications, higher compression rates are prevalent, leading to substantial information loss.

\begin{figure}
	\centering
		\includegraphics[scale=1, width=0.85\linewidth]{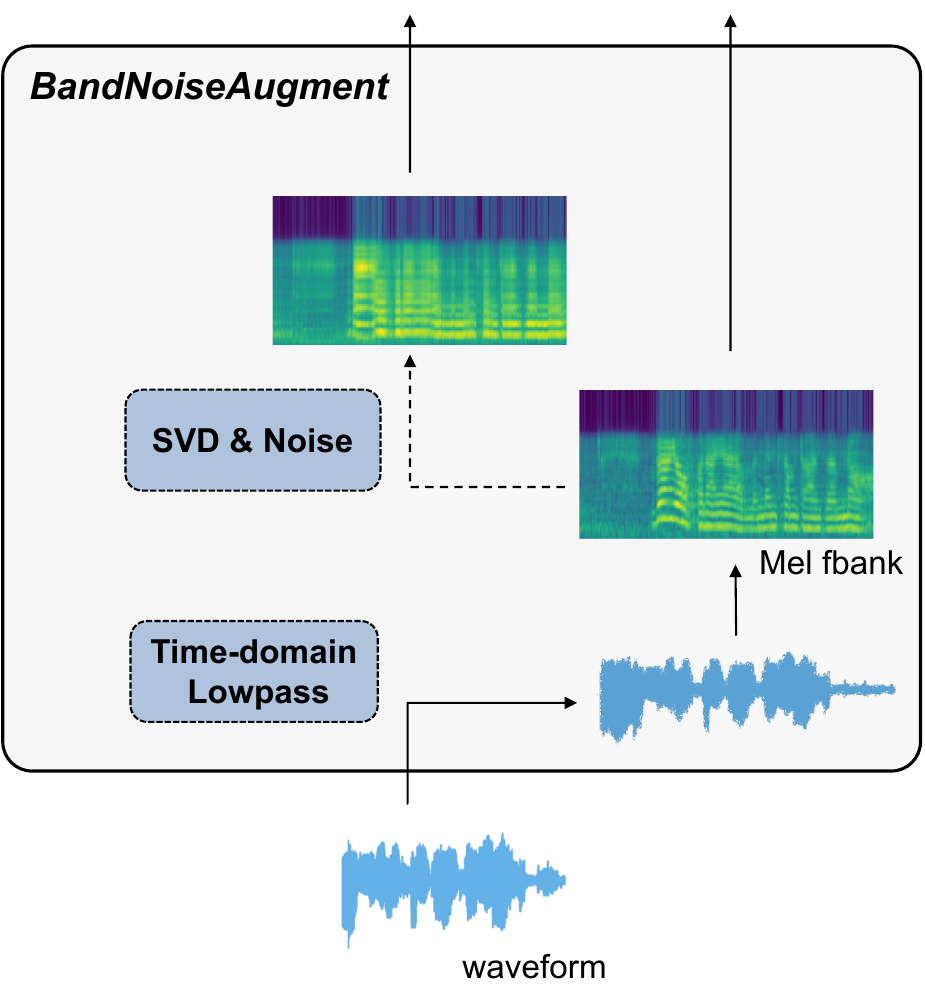}
	\caption{BandNoiseAugment Module. Bandwidth manipulation is applied to the waveform. SVD and noise injection are applied to Mel fbanks.}
	\label{FIG:noiseaug}
\end{figure}

To bridge this gap, we propose the utilization of \textbf{bandwidth manipulation}. This approach systematically incorporates the variability in compression rates encountered in diverse communicative contexts, thereby enhancing the robustness and generalizability of speaker verification systems. For time-domain speech signals with low sampling rates, multiple low-pass filters can be applied directly for time-domain filtering as in Figure.\ref{FIG:noiseaug}. Here we design N-order Butterworth Low-pass Filters for waveforms. Given a sample frequency \(f_s\), cutoff frequency \(f_c\) and order \(N\), then the normalized cutoff frequency \(\omega_c\) is:

\begin{equation}
\omega_c = \frac{f_c}{f_s}
\label{eq1}
\end{equation}

The magnitude response \( |H(\omega_c)| \) of an N-order Butterworth filter is given by:

\begin{equation}
|H(\omega_c)| = \frac{1}{\sqrt{1+(\frac{\omega}{\omega_c})^{2N}}}
\label{eq2}
\end{equation}

where \(\omega\) is the angular frequency. For implementation, the analog Butterworth filter is transformed into a digital filter using the bilinear transform. The transfer function of the digital Butterworth filter in the z-domain is typically expressed in terms of its second-order sections (SOS) for numerical stability. For each second-order section, the transfer function \(H_{sos}(z)\) can be written as:

\begin{equation}
H_{sos}(z) = \frac{b_0 + b_1z^{-1} + b_2z^{-2}}{1 + a_1z^{-1} + a_2z^{-2}}
\label{eq3}
\end{equation}

where \(b_0, b_1, b2\) are the feedforward coefficients, and \(a_1, a_2 \)are the feedback coefficients. For an N-order Butterworth filter, the overall transfer function \(H(z)\) is the product of the transfer functions of each second-order section:

\begin{equation}
H(z) = \prod\limits_{k=1}\limits^{\lceil N/2 \rceil} H_{sos_k}(z)
\label{eq4}
\end{equation}

Then, the filtering operation for a discrete-time audio signal \(x[n]\) using the second-order sections is performed sequentially. For each SOS, the output \(y[n]\) is computed as:

\begin{align}\label{eq5}
y[n] = b_0x[n] &+ b_1x[n-1] + b_2x[n-2]\\
 &- a_1x[n-1] - a_2x[n-2]
\end{align}


For acoustic features in frequency domains, both SpecAugment and Random Erase methodologies draw inspiration from the fundamental concept of dropout. They operate by applying a masking matrix to the input acoustic features, intentionally omitting certain information to emulate signal loss or corruption. For a given \(t \times f\) matrix \(X\) (audio features), the augmented \(X'\) is:

\begin{equation}
X' = X \cdot M
\label{eq6}
\end{equation}

Where \(M\) denotes a binary mask matrix of identical shape as \(X\), comprised of elements that are exclusively 0 or 1. For dropout, each element in \(M\) is sampled from a Bernoulli distribution. For SpecAugment, the process commences with the selection of a random masking length \(L\) drawn from the Uniform distribution. Thereafter, a segment of length \(L\) starting from a randomly chosen temporal or frequency dimension is identified within the mask matrix, with all corresponding elements within this segment being set to zero.

Inspired by these methods, we propose the utilization of \textbf{Noise Injection}, which injects Gaussian noise as a generalized form of Dropout \cite{10.5555/2627435.2670313}. Considering the differences in speech data between scenarios such as radio communications and general environments—where signals are often compressed—we propose data augmentation in the compressed speech feature space. The compressed space here is constructed through the singular value decomposition (SVD) as in Figure.\ref{FIG:noiseaug}. SVD decomposes a matrix into three constituent matrices. Specifically, SVD represents \(X\) as the product of three matrices:

\begin{equation}
X = U \varSigma V^T
\label{eq7}
\end{equation}

Data compression using SVD involves approximating the original matrix with a lower-rank matrix, which retains the most significant features while reducing the size of the data. The chosen rank \(k\) (where \(k < min(t,f)\)), which determines the level of approximation and compression. The compressed approximation \(X_k\) of the original matrix \(X\) is given by:

\begin{equation}
X_k = U_k \varSigma_k V_k^T
\label{eq8}
\end{equation}

Here we inject Gaussian noise into the matrix \(V\), as in Fig.\ref{FIG:noiseaug}. So the compressed and augmented \(X'_k\) is computed by:

\begin{equation}
    X'_k = U_k \varSigma_k V_{'k}^{T} = U_k \varSigma_k ((J + \epsilon ) \odot V_{k}^{T})
\label{eq9}
\end{equation}

where \(\epsilon \) is sampled from the Gaussian distribution and \(J\) is the all-ones matrix with the same shape as \(V_k^T\).

\subsection{Early Fine-tuning} 

\begin{figure}[hbt]
\centering
    \includegraphics[scale=1, width=0.65\linewidth]{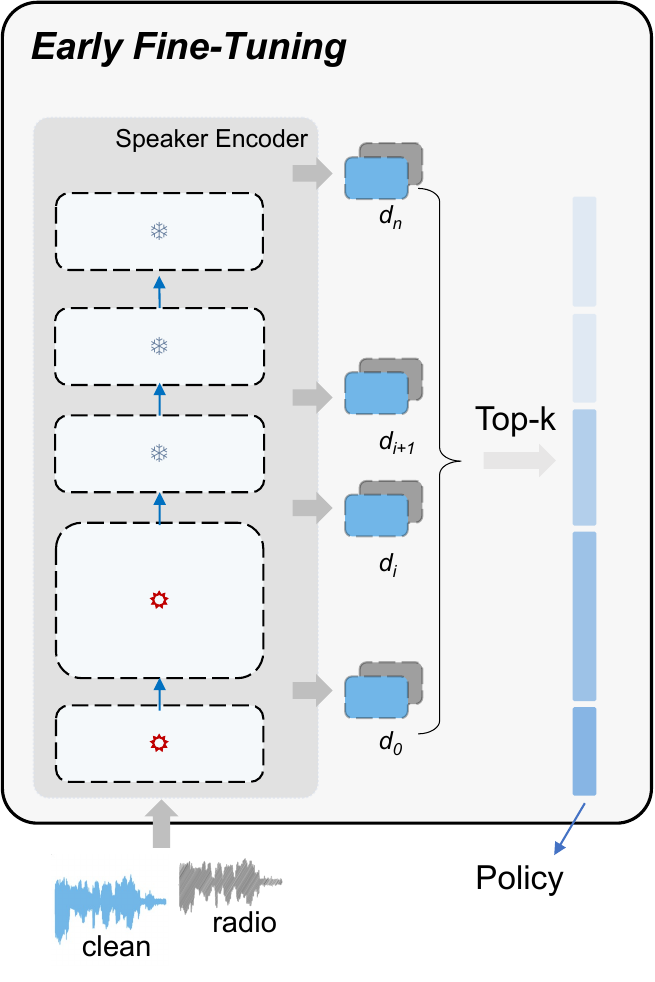}
\caption{Early fine-tuning for radio speaker classification. The fine-tuning policy is derived from comparing the statistical drift (\(d_i\) in Equation.\ref{eq8}) of output features between clean and radio corpora.}
\label{FIG:earlyfine}
\end{figure}

Given that alterations in speech signals, whether due to bandwidth limitations or additional noise, tend to more prominently affect lower-level features while leaving higher-level semantic information relatively intact, networks tasked with extracting such features must exhibit heightened robustness when confronted with low-bandwidth or noisy speech signals. This necessitates a particular emphasis on the resilience of network components dedicated to extracting low-scale features. 

Capitalizing on this characteristic, we propose a fine-tuning policy selection called Early Fine-tuning, specifically for the shallow convolutional layers, thereby enhancing their capability to handle these challenging signal conditions. Our methodology involves feeding two categories of speech corpora into neural networks, extracting features from each layer, and subsequently calculating the Wasserstein distance between the feature statistics of the radio and clean speech datasets. 

\begin{equation}
    d_{i} = Wasse(f_{i}(x_{radio}), f_{i}(x_{clean}) )
    \label{eq10}
\end{equation}

where \(f_{i}(x)\) is the output of the \(i\)-th layer in the classification model. The Wasserstein distance is chosen for its suitability in computing distances between distributions. Layers with larger distance values and greater statistical drift in the speaker classification model are selected for fine-tuning. More precisely, we fine-tune the convolutional layers and utilize larger learning rates for the shallow layers, as in Fig.\ref{FIG:earlyfine}, enhancing the robustness in early-stage feature extraction.  This fine-tuning method for radio speaker verification diverges from conventional practices that often entail fine-tuning the entire embedding extractor or the backend classifier. 

In summary, the overall algorithm for our CRSL framework is listed in Algorithm.1.

\begin{figure}[!t]
    \label{alg:early}
    \renewcommand{\algorithmicrequire}{\textbf{Input:}}
    \renewcommand{\algorithmicensure}{\textbf{Output:}}
    \begin{algorithm}[H]
        \caption{The algorithm of proposed CRSL.}
        \begin{algorithmic}[1]
            \REQUIRE {$D$ clean corpus, $D'$ corrupted corpus};
            \REQUIRE $M$ initial classification model;
            \REQUIRE $p_0 \in [0, 1]$ for augmentation probability, $f_c \in [0, f_s/2]$, $\lambda$ for noise energy;
            \REQUIRE $N_0$ \#minibatch of clean corpus, $N_1$ \#minibatch of fine-tuning corpus;
            \FOR{ $i=0$ to $N_0$}
                \STATE $p_i \sim Uniform(0, 1)$
                \IF{$p_i \leq p_0$}
                    \STATE $x_{N_0} = Bandwidth(x_{N_0}, f_c)$
                    \STATE $feat_{N_0} = AcousticFeaturs(x_{N_0})$
                    \STATE $feat_{N_0} = NoiseInjection(feat_{N_0}, \lambda)$
                \ELSE
                    \STATE $feat_{N_0} = AcousticFeaturs(x_{N_0})$
                \ENDIF
                \STATE Optimize M using minibatch $feat_{N_0}$;
            \ENDFOR
            \STATE Save model $M_0$;
            \STATE 
            \STATE Compute $[d_{0}, d_{1}, ..., d_{n}]$ with Equation.\ref{eq10} and select top-k layers;
            \FOR{ $i=0$ to $N_1$}
                \STATE $p_i \sim Uniform(0, 1)$
                \IF{$p_i \leq p_0$}
                    \STATE $x_{N_1} = Bandwidth(x_{N_1}, f_c)$
                \ENDIF
                \STATE $feat_{N_1} = AcousticFeaturs(x_{N_1})$
                \STATE Optimize M with minibatch $feat_{N_1}$;
            \ENDFOR
            \STATE \RETURN Model $M_0$, Fine-tuned Model $M_1$
        \end{algorithmic}
    \end{algorithm}
\end{figure}

\section{Experiments}
\label{sec:experiments}

\subsection{Dataset}

\noindent \textbf{VoxCeleb}\cite{NAGRANI2020101027}: This large-scale English audio speaker dataset includes VoxCeleb1 and VoxCeleb2. VoxCeleb1 has 1,211 speakers with 148,642 utterances for training and 40 speakers with 4,870 utterances for testing. The VoxCeleb2 development set, used for training, contains 5,994 speakers with 1,092,009 utterances, totaling over 2,700 hours of audio.

\noindent \textbf{CNCeleb}\cite{fan2020cn}: This large-scale Chinese audio speaker dataset spans multiple genres and includes 3,000 speakers with 632,740 utterances for training and 17,973 utterances from 200 speakers for testing, totaling over 1,300 hours of audio.

We use our corpus collection toolkit to obtain VoxCeleb1-A (test) from pipeline A, and VoxCeleb1-B, VoxCeleb2-B, and CNCeleb-B from pipeline B.

\subsection{Training Details}

We implement the ECAPA-TDNN \cite{desplanques2020ecapa} deep speaker verification models within the PyTorch framework. These models are trained on VoxCeleb2, VoxCeleb1, and CNCeleb datasets. For VoxCeleb1, the number of channels is set to 256, while VoxCeleb2 and CNCeleb use 512 channels. 

For model training, we employ the Adam optimizer in PyTorch, setting the weight decay to 0.00002 for extractor layers and 0.0002 for classifier layers, with a momentum of 0.9. Input features consist of 80-dimensional Mel-banks extracted from 2-second segments. The batch size is fixed at 96. Data augmentation follows the method employed in the training of \textit{x-vectors} \cite{snyder2018x}, utilizing the MUSAN \cite{snyder2015musan} dataset.

For the loss function, we adopt the Additive Angular Margin Softmax (AAM-Softmax) \cite{deng2019arcface}, with a scale \(s\) of 30 and a margin \(m\) of 0.2.

During testing, cosine similarity is computed without score normalization. Model performance is evaluated using the Equal Error Rate (EER) and Minimum Detection Cost Function (minDCF) with \(P_{target} =0.01\). Lower values of both EER and minDCF indicate superior model performance. The test trials encompass three subsets for VoxCeleb1: Original (Vox1-O), Easy (Vox1-E), and Hard (Vox1-H). The test trials file of CNCeleb is the official one. We named the Vox1-O trials from the narrowband and wideband datasets of pipeline A as \textbf{Vox1-NH} and \textbf{Vox1-WH}, respectively. The Vox1-O trials from the narrowband and wideband datasets of pipeline B are named \textbf{Vox1-N} and \textbf{Vox1-W}, respectively.

\subsection{Hyperparameters Analysis}
\label{sec:speech_analysis}
In light of the tunable hyperparameters within both the CRSL framework's BandNoiseAugment and the Early Fine-tuning Module discussed in Section \ref{sec:meth}, we now provide insight into the rationale behind the selected hyperparameter configurations, drawing from a comprehensive analysis of our dataset's characteristics.

\subsubsection{Analysis for BandNoiseAugment Module}

\begin{figure}[h]
	\centering
 \subfloat[NBFM]{
		\includegraphics[scale=.95, width=0.85\linewidth]{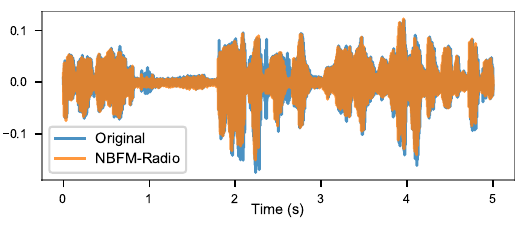}
  }
  \hspace{0pt}
 \subfloat[WBFM]{
	\includegraphics[scale=.95, width=0.85\linewidth]{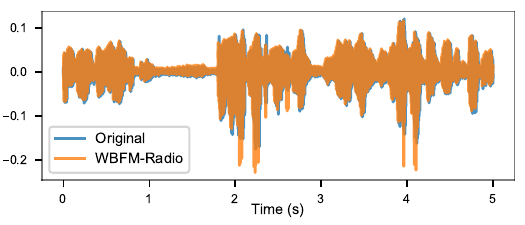}
 }
\caption{Comparison of waveforms from NBFM and WBFM with original audio. The loss of detail is evident in the NBFM radio audio. The blue waveform is the original audio. The orange waveform is radio audio.}
\label{FIG:audio_waveform}
\end{figure}

In our comparative assessment, we analyze the discrepancies between radio audio samples derived from \textbf{Pipeline A} and the pristine originals across both time and frequency domains. The investigation reveals that the most notable distinction pertains to bandwidth. To provide concrete evidence of this difference, we choose a specific audio sample from VoxCeleb1, namely (id10270/5r0dWxy17C8/00001.wav), and examine its waveform and log spectrogram over a 5-second segment. This comparative analysis vividly demonstrates the effects of radio transmission on the audio sample's characteristics.


\begin{figure}[h]
    \centering
    \subfloat[Original]{ 
        \includegraphics[scale=1, width=0.85\linewidth]{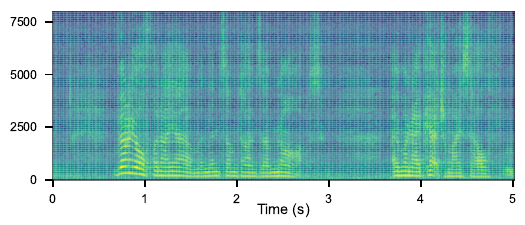}}
    \hspace{0pt}
    \subfloat[NBFM]{ 
        \includegraphics[scale=1, width=0.85\linewidth]{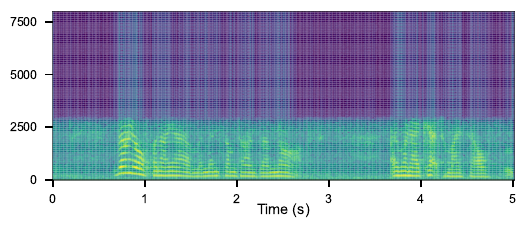}
        }
    \hspace{0pt}
    \subfloat[WBFM]{ 
        \includegraphics[scale=1, width=0.85\linewidth]{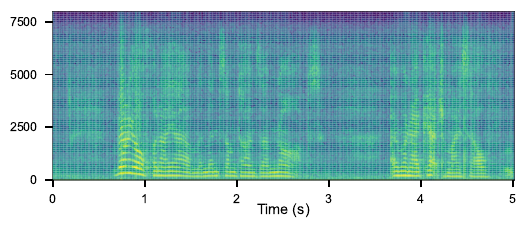}}
    \caption{Comparison of Log Spectrograms from Original, NBFM, and WBFM audios from Pipeline A. Details for frequencies roughly above 3000 Hz in the NBFM transmission are absent.}
\label{FIG:spectrograms}
\end{figure}

Diverse bandwidths significantly impact the quality of transmitted speech signals in analog radio transmissions using GNU Radio and HackRF One. By default, the signal passes through distinct low-pass filters based on the bandwidth type. The NBFM receive block applies a filter with a 2700 Hz high cutoff frequency, whereas the WBFM Transmit block uses a filter with a 16000 Hz high cutoff frequency.



In Figure.\ref{FIG:audio_waveform} and Figure.\ref{FIG:spectrograms}, comparing the original audio to the NBFM radio audio reveals clear differences. In the time domain, the radio audio exhibits a noticeable loss of detail, with the WBFM sample demonstrating less loss compared to the NBFM. Moreover, the WBFM audio's signal amplitude notably surges above that of the original signal at approximately 2s and 4s.

In the frequency domain, disparities become far more conspicuous. The NBFM signal undergoes a low-pass filtering process, resulting in the attenuation of high-frequency information; thus, spectrogram details for frequencies roughly above 3000 Hz in the NBFM transmission are absent. Similarly, for WBFM samples, signal loss occurs in the frequency range exceeding 7000 Hz.

In summary, we set the cutoff frequencies (\(f_c\)) of the BandNoiseAugment module to [2000, 3000, 5000, 7000] Hz. During data augmentation, one of these cutoff frequencies is randomly selected. 

\subsubsection{Analysis for Early Fine-Tuning}
\label{subsec:feature}

This section aims to analyze the differential effects of neural network-based speaker vector extraction models when applied to radio speech corpus versus clean data. Through feature divergence analysis, we investigate the statistical shifts in feature mappings across different domains. Our methodology involves feeding two categories of speech corpora into neural networks, extracting features from each layer, and subsequently calculating the Wasserstein Distance between the feature statistics of the radio and clean speech datasets. 

\begin{figure}[!tb]
	\centering
    \subfloat[Trainset: VoxCeleb1]{ 
        \includegraphics[scale=1, width=0.85\linewidth]{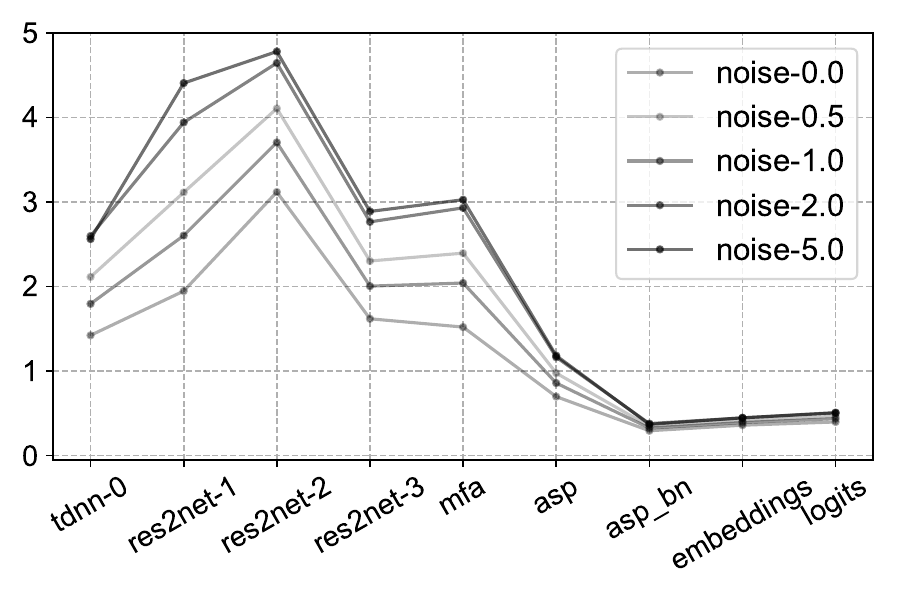}}
    \hspace{0pt}
    \subfloat[Trainset: VoxCeleb1, VoxCeleb1-N]{ 
        \includegraphics[scale=1, width=0.85\linewidth]{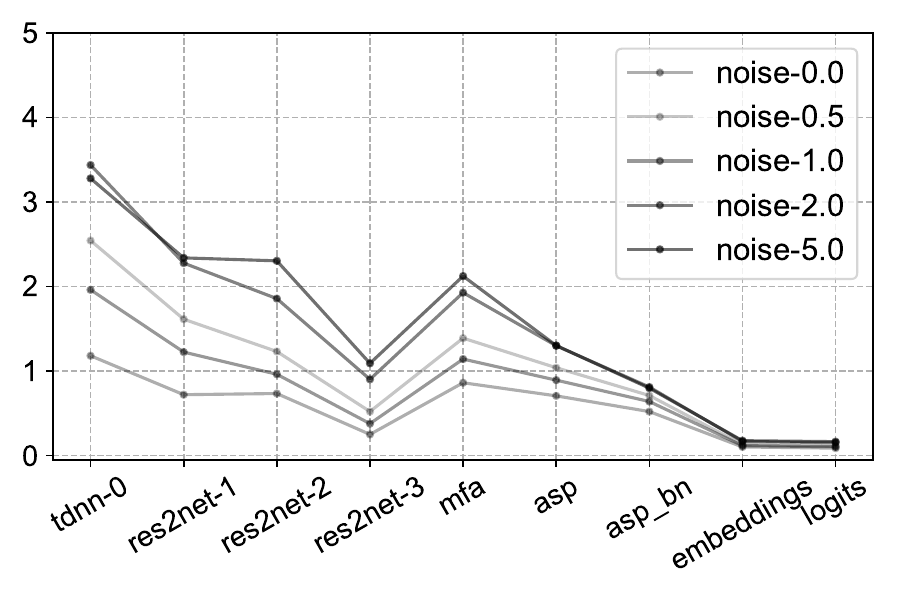}}
    \caption{The comparison of relative feature drift analysis for the ECAPA-TDNN model trained using clean data and the one trained using a combination of clean and radio NBFM speech data. The noise value is the energy of the noise level in pipeline B.}
	\label{FIG:mix_clean}
\end{figure}

Additionally, we trained a mixed model using a combination of radio and clean speech corpora from VoxCeleb1. The visualization of the statistical shifts in these features is depicted in Figure.\ref{FIG:mix_clean}. From Figure.\ref{FIG:mix_clean} (a), it is evident that the difference in features between radio and clean data exhibits a notably larger discrepancy at the convolutional layers compared to the baseline (depicted by the blue curve). There is also a distinct margin between the speaker vectors extracted from the two types of data.

As illustrated in Figure.\ref{FIG:mix_clean} (b), if both types of data are utilized as part of the training set, the differences in the convolutional layer features can be effectively diminished. Consequently, we can conclude that the signal transformation (or degradation) introduced by the radio transmission in audio interferes with the speaker feature extraction capabilities of the existing ECAPA-TDNN model, with a relatively pronounced impact on its convolutional layers.

\subsection{Results}
  
In this section, we first present the large-scale radio speaker verification benchmark on VoxCeleb1\&2 and the performance of the proposed CRSL framework. Subsequently, we delve into the reasons behind the performance degradation observed in the radio corpus and show the hyperparameter of sub-modules in our framework. Lastly, we carry out ablation studies.
  
\subsubsection{Basic Results}

\begin{table*}[h]
  \scriptsize
  \centering
  \caption{Overall results on the VoxCeleb1. The noise level for the radio VoxCeleb1 corpus is 1.0. \textbf{Vox1-N}: the test set is NBFM corpus from pipeline B. \textbf{Vox1-W}: the test set is WBFM corpus from pipeline B. \textbf{Bold} font highlights the best performance.}
  \label{tab:overall}
  \tabcolsep=4pt
  \renewcommand\arraystretch{1.1}
\begin{tabular}{ccccccc}
\toprule
\multirow{2}{*}{\textbf{Method}}                               & \multicolumn{2}{c}{\textbf{Vox1-O}} & \multicolumn{2}{c}{\textbf{Vox1-N}} & \multicolumn{2}{c}{\textbf{Vox1-W}} \\ \cmidrule{2-7}
                                                               & \multicolumn{1}{c}{EER(\%)$\downarrow$} & \multicolumn{1}{c}{minDCF$\downarrow$}         & \multicolumn{1}{c}{EER(\%)$\downarrow$} & \multicolumn{1}{c}{minDCF$\downarrow$}          & \multicolumn{1}{c}{EER(\%)$\downarrow$} & \multicolumn{1}{c}{minDCF$\downarrow$}          \\ \midrule
-                                                              & \textbf{1.19}         & 0.1410         & 8.99          & 0.7540          &     1.75          &  0.1892               \\
SpechAug \cite{wang2020investigation}         & 1.26         & 0.1230         & 8.96          & 0.7153          &       1.82        &    0.1785             \\
inver\_SpechAug \cite{9414676} & 1.33         & \textbf{0.1287}         & 8.03          & 0.6787          &  1.84             &            0.1682     \\
dropout \cite{10.5555/2627435.2670313}        &  1.22            &   0.1247             &  8.30             &      0.7292           &   1.89            & \textbf{0.1641}                \\
CRSL (Ours)                                                    & 1.21         & 0.1431         & \textbf{4.32}          & \textbf{0.4414}          &    \textbf{1.64}           &   0.1827             \\
\bottomrule
    \end{tabular}
  \end{table*}
  
The foundational test results are presented in Table.\ref{tab:overall}. We contrast our framework with prior data augmentation approaches on radio corpora. It is evident that the proposed CRSL framework attains the lowest EER among radio corpora, thus enhancing speaker verification models, particularly for NBFM radio communication.

\subsubsection{Large-scale Radio Speaker Benchmark}
\begin{table*}[h]
  \scriptsize
  \centering
  \tabcolsep=4pt
  \renewcommand\arraystretch{1.1}
  \caption{Results of various models on the Original VoxCeleb1 test trials. An asterisk (*) signifies pre-trained models obtained from the wespeaker toolkit \cite{wang2023wespeaker}. \textbf{Vox1-NH}: the test set is NBFM corpus from pipeline A. \textbf{Vox1-WH}: the test set is WBFM corpus from pipeline A.}
  \label{tab:pretrain}
\begin{tabular}{lcrrrrrr}
\toprule
\multicolumn{1}{c}{\multirow{2}{*}{\textbf{System}}}     & \multirow{2}{*}{\textbf{\#Params (M)}} & \multicolumn{2}{c}{\textbf{Vox1-O}}                                               & \multicolumn{2}{c}{\textbf{Vox1-NH}}                                                & \multicolumn{2}{c}{\textbf{Vox1-WH}}                                                \\
\cmidrule{3-8}
\multicolumn{1}{c}{}                                     &                                  & \multicolumn{1}{c}{EER(\%)$\downarrow$} & \multicolumn{1}{c}{minDCF$\downarrow$} & \multicolumn{1}{c}{EER(\%)$\downarrow$} & \multicolumn{1}{c}{minDCF$\downarrow$} & \multicolumn{1}{c}{EER(\%)$\downarrow$} & \multicolumn{1}{c}{minDCF$\downarrow$} \\
\midrule
ECAPA-TDNN* \cite{desplanques2020ecapa} & 6.19                             & 1.17                                    & 0.1264                                 & 5.02                                    & 0.5153                                 & 3.42                                    & 0.2317                                 \\
ResNet34* \cite{xie2019utterance}       & 6.63                             & 1.07                                    & 0.0996                                 & 3.38                                    & 0.4131                                 & 3.00                                    & 0.2117                                 \\
CAM++* \cite{wangcam++}                 & 7.18                            & 0.92                                    & 0.1104                                 & 3.20                                    & 0.3793                                 & 2.72                                    & 0.2018                                 \\
\midrule
ECAPA-TDNN                              & 6.19                           & 1.15                                    & 0.1385                                 & 4.98                                    & 0.5308                                 & 3.04                                    & 0.2666                         \\       
\bottomrule
  \end{tabular}
  \end{table*}
  
To establish a large-scale benchmark for radio speaker verification, we initially assess pre-trained models using the radio speech corpus captured by the HackRF One device (referred to as Pipeline A). As shown in Table.\ref{tab:pretrain}, our implemented ECAPA-TDNN yields results comparable to those of the model from the Wespeaker toolkit \cite{wang2023wespeaker}. The close proximity between the receiver and transmitter ensures superior signal quality. However, all models exhibit markedly poorer performance on the radio corpus compared to clean speech data. Interestingly, ResNet34 and CAM++ display a smaller decline than ECAPA-TDNN. We speculate that the 2D-convolutional model appears to be more robust than the 1D-convolutional model in radio scenarios.

Specifically, for the ECAPA-TDNN, the EER increases by approximately 1.8\% on the WBFM corpus and by about 3.8\% on the NBFM corpus. Moreover, performance on the NBFM test set is notably inferior to that on the WBFM test dataset. The primary distinction between WBFM and NBFM corpora lies in their bandwidth. Thus, we deduce that bandwidth limitations inherent to narrowband and wideband transmission contribute to nearly 50\% of the performance degradation, with the remaining 50\% attributed to other channel-related factors.

To gather a large-scale radio speech corpus and establish a benchmark for radio speaker verification, we simulated the transmission of the VoxCeleb1 dataset across diverse channels for evaluation purposes using \textbf{Pipeline B}. Our radio speech data from the VoxCeleb1 dataset was transmitted under NBFM conditions, with noise levels ranging from 0 to 5.0. We compared our results with previous radio corpora to assess the noise levels in our corpus, as presented in Table.\ref{tab:vox1_noiselevel}.

\begin{table}[h]
  \centering
  \scriptsize
  \tabcolsep=2pt
  \renewcommand\arraystretch{1.1}
  \caption{Results of previous studies and results on radio corpus. VoxCeleb1-A is from pipeline A and VoxCeleb1-B is from pipeline B.}
  \label{tab:vox1_noiselevel}
    \begin{tabular}{llcrr}
      \toprule
\multicolumn{1}{c}{\textbf{System}}                                    & \textbf{Testset}                            & \textbf{Noise Level} & \multicolumn{1}{c}{\textbf{EER} (\%)$\downarrow$} & \textbf{minDCF}$\downarrow$ \\
 \midrule
        x-vector \cite{trnka2021speaker}            & VoxCeleb1-Radio                     & -           & 5.9                                      & -                  \\
	i-vector \cite{plchot2013developing}        & RATS SID 30s   & -           & 5.3                                      & -                  \\
	i-vector \cite{McLaren2014ApplicationOC}    & RATS SID 10s                       & -           & 8.1                                      & -                  \\
	CNN-ivector \cite{McLaren2014ApplicationOC} & RATS SID 10s                       & -           & 6.5                                      & -                  \\
	ivector+WCC \cite{glembek2014domain}        & RATS SID                           & -           & 6.26                                     & -                  \\
	UBM-GMM-HMM \cite{larcher2014extended}      & \multicolumn{1}{l}{RSR2015 M (F)}   & -           & 5.79 (4.40)                                    & -                  \\
\midrule
\multirow{7}{*}{ECAPA-TDNN (Ours)}                                  & \multirow{1}{*}{VoxCeleb1}    & -           & 1.14                                     & 0.1385             \\ 
                                                             &  \multirow{1}{*}{VoxCeleb1-A}                                  & -         & 4.98                                     & 0.5308             \\ \cmidrule{2-5}
                                                             &  \multirow{5}{*}{VoxCeleb1-B}                               & 0.0         & 4.10                                     & 0.4621             \\
                                                             &                                    & 0.5         & 5.89                                     & 0.5670             \\
                                                             &                                    & \underline{1.0}         & 7.34                                     & 0.6351             \\
                                                             &                                    & 2.0         & 10.79                                    & 0.8394             \\
                                                             &                                    & 5.0         & 12.04                                    & 0.9194             \\
      \bottomrule
      \end{tabular}
\end{table}

The primary distinction among these channels lies in the noise energy present in their respective channel models. The results indicate a sharp deterioration in the model's performance with increasing noise energy. Simulated channel data with noise levels ranging from 0.5 to 1.0 exhibit performances akin to those derived from actual hardware transmissions using the HackRF One device. Therefore, we predominantly utilized VoxCeleb1 data with a noise level of 1.0 for primary testing, recognizing the challenge it poses for speaker verification. Subsequently, we simulated the transmission of the VoxCeleb2 dataset under a channel with a noise level of 1.0, resulting in the creation of a radio corpus for VoxCeleb2, which we utilized for further training.

\begin{table*}[h]
  \centering
  \tabcolsep=6pt
  \renewcommand\arraystretch{1.1}
    \caption{Results on the clean and radio speech data from the VoxCeleb1 and CNCeleb dataset transmitted under NBFM channel, with noise levels is 1.0. VoxCeleb2-B and CnCeleb-B are from pipeline B.}
  \label{tab:vox_cnc}
\begin{tabular}{lcccccc}
\toprule
\multirow{2}{*}{\textbf{Trainset}} & \multirow{2}{*}{\textbf{Duration} (hours)} & \multicolumn{2}{c}{\textbf{Trials-O}}                                                 & \multicolumn{2}{c}{\textbf{Trials-N}}                                               \\ \cmidrule{3-6}
                                   &                                    & \multicolumn{1}{c}{EER (\%)$\downarrow$} & \multicolumn{1}{c}{minDCF$\downarrow$} & \multicolumn{1}{c}{EER (\%)$\downarrow$} & \multicolumn{1}{c}{minDCF$\downarrow$} \\ \midrule
VoxCeleb2                               & 1 \(\times\) 2354                                 & 1.15                                     & 0.1385                                 & 7.34                                     & 0.6351                                 \\
VoxCeleb2+VoxCeleb2-B                      & 2 \(\times\) 2354                                 & 1.53                                     & 0.2126                                 & 3.79                                     & 0.4434                                 \\ \midrule
CnCeleb                                & 1 \(\times\) 1312                                & 8.32                                     & 0.4460                                 & 22.91                                    & 0.9998                                 \\
 CnCeleb+CnCeleb-B                       & 2 \(\times\) 1312                                & 10.23                                    & 0.5417                                 & 12.71                                    & 0.6359                                 \\ 
      \bottomrule
\end{tabular}
\end{table*}

Subsequently, we simulated the transmission of the VoxCeleb2 and CNCeleb datasets, subsequently mixing the radio speech corpus with their respective clean counterparts for training speaker verification models. The outcomes of this mixed training regimen are summarized in Table.\ref{tab:vox_cnc}. \textbf{Duration} column gives the total duration of speech in the training sets. Models trained on the combined data exhibit substantial improvement when evaluated on radio speech test sets. Specifically, for VoxCeleb, the EER on radio data testing experienced an absolute reduction of 3.55\%, while in CNCeleb, this absolute decrease reached 10.2\%. Notably, however, all these models displayed marked performance declines on their original test sets.

\subsubsection{BandNoiseAugment}
\begin{table}[!bh]
  \centering
  \scriptsize
  \tabcolsep=6pt
  \renewcommand\arraystretch{1.1}
    \caption{Results on the clean and radio speech data for models with types of augment methods.}
  \label{tab:vox_band}
\begin{tabular}{lcccc}
\toprule
\multicolumn{1}{c}{\multirow{2}{*}{\textbf{Augment}}}              & \multicolumn{2}{c}{\textbf{Vox1-O}}                                                & \multicolumn{2}{c}{\textbf{Vox1-N}}                                                 \\
\cmidrule{2-5}
\multicolumn{1}{c}{}                                               & \multicolumn{1}{c}{EER (\%)$\downarrow$} & \multicolumn{1}{c}{minDCF$\downarrow$} & \multicolumn{1}{c}{EER (\%)$\downarrow$} & \multicolumn{1}{c}{minDCF$\downarrow$} \\
\midrule
-                                 & 1.36                                     & 0.1628                                 & 16.11                                     & 0.8918                                 \\
+env\_crop \cite{ravanelli2021speechbrain}                                 & 1.18                                     & 0.1419                                 & 8.80                                     & 0.7540                                 \\
+spechaug \cite{wang2020investigation}                                  & 1.26                                     & 0.1230                                 & 8.96                                     & 0.7153                                 \\
+inver\_spechaug \cite{9414676}                           & 1.33                   & 0.1287                   & 8.03                   & 0.6787                   \\ 
\midrule
+noise                                                         & 1.24                                     & 0.1420                                 & 8.66                                     & 0.7089                                 \\
+band                                                         & \textbf{1.19}                                    & \textbf{0.1241}                                 & 5.31                                     & 0.5514                                 \\
+band +noise                                                        & 1.24                                     & 0.1449                                 & \textbf{5.08}                                     & \textbf{0.5174}                                \\
      \bottomrule
\end{tabular}
\end{table}

ECAPA-TDNN with previous data augmentation methods shows a significant improvement in performance over the ECAPA-TDNN without data augmentation when evaluated on radio speech corpora. This result suggests that common data augmentation methods, including additive noise, reverberation, and time-frequency masking, can enhance the robustness of speaker verification models in radio environments. However, the relative degradation between clean, NBFM, and WBFM corpora remains consistent. This implies that the data augmentation methods used in our experiment do not specifically address the performance degradation caused by bandwidth and other channel factors. Therefore, it may be necessary to devise tailored data augmentation strategies for speaker verification in radio scenarios.

We then validate the BandNoiseAugment module on the simulated VoxCeleb1 datasets, where only clean VoxCeleb2 dev data is used for training. The outcomes are summarized in Table.\ref{tab:vox_band}. Results evidence that among prior channel bandwidth-oriented data augmentation techniques, inver\_spechaug \cite{9414676} proves most efficacious; however, these methods risk degradation on the original dataset with the same training configuration. 

Conversely, our proposed BandNoiseAugment exhibits minimal performance degradation on the baseline test set and marked improvement on radio-transmitted datasets, illustrating its superior efficacy. Similarly, the effectiveness of each submodule within BandNoiseAugment is also corroborated by the data presented in this table. Specifically, for VoxCeleb, the EER on radio data testing experienced an absolute reduction of 3.72\% for our method.

\subsubsection{Early Fine-tuning}
\label{subsec:fine}

\begin{table}[h]
\centering
  \scriptsize
 \tabcolsep=4pt
  \renewcommand\arraystretch{1.1}
    \caption{Fine-tuning results on the clean and NBFM radio VoxCeleb1 test.}
  \label{tab:fine}
  \begin{tabular}{ccccccc}
      \toprule
\multirow{2}{*}{\textbf{Layers}}   & \multirow{2}{*}{\textbf{\#Param(M)}} & \multirow{2}{*}{\textbf{LR}} & \multicolumn{2}{c}{\textbf{Vox1-O}}                                                & \multicolumn{2}{c}{\textbf{Vox1-N}}                                                 \\
\cmidrule{4-7}
                                   &                                      &                              & \multicolumn{1}{c}{EER (\%)$\downarrow$} & \multicolumn{1}{c}{minDCF$\downarrow$} & \multicolumn{1}{c}{EER (\%)$\downarrow$} & \multicolumn{1}{c}{minDCF$\downarrow$} \\ 

 \midrule
Full                               & 7.34 (100\%)                         &     -                         & 1.19                                     & 0.1464                                 & 5.07                                     & \textbf{0.4863}                                 \\
Full  & 7.34 (100\%)                         &     $\Uparrow$                         & 1.26                                     & 0.1619                                 & 5.43                                     & 0.4940                                 \\
Full  & 7.34 (100\%)                         &    $\Downarrow$                           & \textbf{1.17}                                     & \textbf{0.1363}                                 & \textbf{4.77}                                     & 0.4866                                 \\ \midrule
0123                               & 2.45 (33.3\%)                        & -                            & 1.20                                     & 0.1371                                 & 5.22                                     & \textbf{0.4885}                                 \\
1234                               & 4.60 (62.7\%)                        & -                            & \textbf{1.18}                                     & \textbf{0.1367}                                 & \textbf{5.12}                                     & \textbf{0.4885}                                 \\
2345                               & 4.65 (63.3\%)                        & -                            & 1.21                                     & 0.1373                                 & 5.28                                     & 0.5014                                 \\
3456                               & 3.90 (53.1\%)                        & -                            & 1.26                                     & 0.1406                                 & 5.67                                     & 0.5205                                 \\
4567                               & 3.75 (51.1\%)                        & -                            & 1.23                                     & 0.1360                                 & 5.92                                     & 0.5337                                 \\ 
\midrule
1234                               & 4.60 (62.7\%)                        & $\Uparrow$                   & 1.41                                     & 0.1493                                 & 5.18                                     & 0.4956                                 \\
1234                               & 4.60 (62.7\%)                        & $\Downarrow$                 & \textbf{1.17}                                    & \textbf{0.1356}                                 & \textbf{4.81}                                     & \textbf{0.4790}                                 \\ 
\bottomrule
\end{tabular}
\end{table}

To validate the efficacy of Early Fine-tuning and the insights from feature drift analysis, we fine-tuned different layers of the ECAPA-TDNN model separately and subsequently assessed their performance on both original and radio speech corpora, as presented in Table.\ref{tab:fine}. The \textbf{Layers} column specifies the layers subjected to fine-tuning. The \textbf{Param} column denotes the number of trainable parameters during fine-tuning. The \textbf{LR} column denotes the learning rate strategy employed during fine-tuning, where '-' signifies a constant value of 0.00001, '$\Uparrow$' indicates a learning rate that increases from 0.000001 to 0.001 as the layer depth increases, and '$\Downarrow$' represents a learning rate that decreases from 0.001 to 0.000001 with increasing layer depth.

We chose 16 utterances per speaker from the VoxCeleb2 development set and merged their radio versions with the clean counterparts for fine-tuning, generating close to 415 hours of speech in the fine-tuning set.  We fine-tuned all the models for 4 epochs. The model's layer sequence corresponds to the following structures: tdnn1 (0), SERes2NetBlock1 (1), SERes2NetBlock2 (2), SERes2NetBlock3 (3), MFA (4), Attentive Statistical Pooling (5), ASP\_bn (6), and FC (7). 

\begin{table*}[!th]
  \centering
  \caption{Ablation Study results on three VoxCeleb1 testing trials files. \textbf{Bold} font highlights the best performance.}
  \label{tab:bandpass}
  \tabcolsep=2pt
  \renewcommand\arraystretch{1.1}
  \begin{tabular}{clcccccc}
    \toprule
    \multirow{2}{*}{\textbf{Testset}} & \multicolumn{1}{c}{\multirow{2}{*}{\textbf{Augment}}} & \multicolumn{2}{c}{\textbf{Vox1-O}}                                  & \multicolumn{2}{c}{\textbf{Vox1-E}}                                      & \multicolumn{2}{c}{\textbf{Vox1-H}}                                      \\
                             & \multicolumn{1}{c}{}                         & \multicolumn{1}{c}{EER (\%)$\downarrow$} & \multicolumn{1}{c}{minDCF$\downarrow$} & \multicolumn{1}{c}{EER (\%)$\downarrow$} & \multicolumn{1}{c}{minDCF$\downarrow$} & \multicolumn{1}{c}{EER (\%)$\downarrow$} & \multicolumn{1}{c}{minDCF$\downarrow$} \\ \hline
    \multirow{8}{*}{VoxCeleb1}   & -                                            & 1.36                                  & 0.1628                              & 1.60                                  & 0.1699                              & 2.94                                  & 0.2656                              \\
                             & +aug                                         & \textbf{1.14}                                  & 0.1385                              & 1.39                                  & \textbf{0.1455}                              & 2.51                                  & \textbf{0.2293}                              \\
                             & +bandnoise                                    & 1.48                                  & 0.1771                              & 1.58                                  & 0.1760                              & 2.92                                  & 0.2631                              \\
                             & +earlyfine                                  & 1.40                                  & 0.1620                              & 1.62                                  & 0.1744                              & 2.96                                  & 0.2695                              \\
                             & +aug +bandnoise                               & 1.17                                  & 0.1433                              & \textbf{1.24}                                  & 0.1511                              & \textbf{2.48}                                  & 0.2386                              \\
                             & +aug +earlyfine                              & 1.13                                  & \textbf{0.1337}                              & 1.38                                  & 0.1460                              & 2.49                                  & 0.2344                              \\
                             & +aug +bandnoise +earlyfine                   & 1.21                                  & 0.1431                              & 1.34                                  & 0.1527                              & 2.47                                  & 0.2374                             \\
    \midrule
                             \multirow{8}{*}{VoxCeleb1-B}    & -                                            & 16.11                                 & 0.8918                              & 16.49                                 & 0.9295                              & 23.94                                 & 0.9627                              \\
                             & +aug                                         & 7.34                                  & 0.6351                              & 7.63                                  & 0.6464                              & 12.85                                 & 0.7861                              \\
                             & +bandnoise                                    & 9.48                                  & 0.7705                              & 9.96                                  & 0.7801                              & 16.06                                 & 0.8875                              \\
                             & +earlyfine                                  & 6.78                                  & 0.5682                              & 7.18                                  & 0.6192                              & 12.00                                 & 0.7525                              \\
                             & +aug +bandnoise                               & 5.06                                  & 0.4878                              & 5.35                                  & 0.5192                              & 9.49                                  & 0.8593                              \\
                             & +aug +earlyfine                              & 5.06                                  & 0.4447                              & 5.21                                  & 0.5067                              & 9.16                                  & 0.6674                              \\
                             & +aug +bandnoise +earlyfine                   &  \textbf{4.32}                                  & \textbf{0.4414}    &  \textbf{4.40}                                  & \textbf{0.4488}    &  \textbf{7.83}                                  & \textbf{ 0.5926}   \\
                             \bottomrule
    \end{tabular}
  \end{table*}

Evidently, fine-tuning only the parameters of the shallow convolutional layers (1234) yields superior results. Additionally, the finding that fine-tuning layers 0123 shows worse performance than fine-tuning layers 1234 aligns with our analysis in Section \ref{subsec:feature}. As the depth of the fine-tuned layers increases, performance declines in radio speech tests, with the EER rising from 5.22\% to 5.92\%. When the fine-tuning parameters are held constant, shallow-layer fine-tuning consistently outperforms deep-layer fine-tuning. This conclusion is corroborated by the analysis presented in Figure.\ref{FIG:mix_clean}. 

For radio data, fine-tuning layers 1234 yields an EER of 5.12\%, which is notably lower than the EER of 5.28\% obtained by fine-tuning layers 2345, despite the latter containing more trainable parameters. Additionally, a strategy of decreasing the fine-tuning learning rate with increasing layer depth outperforms both increasing the learning rate and using a fixed learning rate.

\subsection{Ablation Study}

The outcomes of the ablation experiments are presented in Table.\ref{tab:bandpass}. The noise level for the NBFM radio VoxCeleb1 corpus is 1.0. Symbols in the \textbf{Augment} column denote the following: "+aug" signifies vanilla data augmentation, "+bandnoise" represents BandNoiseAugment module, "+earlyfine" refers to early fine-tuning. 

\paragraph{BandNoiseAugment Evaluation.} Given the NBFM radio bandwidth in our study, the cutoff frequency is set at [2000, 3000, 5000, 7000] Hz. For Early Fine-tuning, we use the same fine-tuning configurations (layers 1234) as in Section \ref{subsec:fine}. The BandNoiseAugment module improves performance under radio conditions without extending the model's training duration, as evidenced by a decrease in EER from 7.34\% to 5.06\%. 

\paragraph{Fine-tuning Analysis.} Fine-tuning helps maintain the model's performance on the original test set, as demonstrated by comparably low EERs of 1.14\% and 1.13\% on the original VoxCeleb1 test trials. The results in the table show that combining BandNoiseAugment with fine-tuning enhances the robustness of existing speaker verification models under radio conditions using a smaller amount of radio speech data, evidenced by a decrease in EER from 7.34\% to 4.32\% on the radio VoxCeleb1 test trials. Although this method does not yet match the performance of models trained with the full mixed dataset, it is noteworthy that combining radio and clean data directly leads to a significant performance drop in the original clean speech context, increasing the EER from 1.14\% to 1.53\% in the previous Table.\ref{tab:vox_cnc}.

It is worth noting that despite applying these two compensation strategies or mixing two types of data, the model's performance on the radio corpus still lags behind that on the clean corpus. There is a marked performance gap between the model's performance on clean speech and narrowband radio speech test sets, potentially exceeding those reported in previous studies \cite{plchot2013developing,trnka2021speaker}. Prior research might have reported comparable performance to the clean corpus without considering transmission conditions and noise levels.

\section{Conclusion}
\label{sec:conclusion}

In conclusion, our study presents the Channel Robust Speaker Learning framework, which significantly improves the robustness of speaker verification models under challenging channel conditions. By leveraging data augmentation through the BandNoiseAugment module and implementing early fine-tuning strategies, we have demonstrated the effectiveness of our approach in enhancing model performance without requiring extensive additional training data. Our experimental results show that the BandNoiseAugment method reduces the Equal Error Rate (EER) by 2.28\% without the need for additional training time or data. Moreover, our early fine-tuning technique, which fine-tunes approximately 60\% of the model parameters, achieves superior performance on both the original and radio-transmitted speech datasets.

Furthermore, we have developed a specialized toolkit for simulating radio transmissions and collecting speech data from radio communications, enabling the creation of a comprehensive speech corpus and establishing a benchmark for speaker verification under radio conditions. Moving forward, our research will focus on exploring the variability of speech signals under different complex radio channels to develop even more effective speaker verification methods. This ongoing investigation aims to bridge the gap between simulated conditions and real-world scenarios, thereby advancing the field of speaker verification in radio communications.

\bibliographystyle{IEEEtran}

\bibliography{my}

\end{document}